\documentclass[preprint,12pt]{elsarticle}
\usepackage{amsmath,amsfonts,amssymb,amsthm}
\usepackage{graphicx}
\usepackage{enumerate}
\usepackage[latin1]{inputenc}
\usepackage{latexsym}
\usepackage{color}
\usepackage{dsfont}
\usepackage[english]{babel}

\newcommand{\1}{\mathds{1}}

\renewcommand{\vec}[1]{\boldsymbol{#1}}
\newcommand{\prv}{\vec{\pi}}
\newcommand{\frt}{\vec{\varphi}}

\def\argmin{\operatornamewithlimits{arg\,min}}

\usepackage{url}
\newtheorem{theorem}{Theorem}
\newtheorem{lemma}[theorem]{Lemma}

\journal{}

\begin{document}

\begin{frontmatter}



\title{PageRank Optimization by Edge Selection}


\author[inst1,inst2]{Bal\'azs Csan\'ad Cs\'aji}
\author[inst3,inst4]{Rapha\"el M. Jungers}
\author[inst4]{Vincent D. Blondel}

%\author{}

\address[inst1]{Department of Electrical and Electronic Engineering, School of Engineering,\\
The University of Melbourne, Australia, {\tt bcsaji@unimelb.edu.au}}
\address[inst2]{Computer and Automation Research Institute, Hungarian Academy of Sciences}
\address[inst3]{Lab.\ for Information and Decision Systems, Massachusetts Institute of Technology}
\address[inst4]{Department of Mathematical Engineering, Universit\'e catholique de Louvain, Belgium, {\tt vincent.blondel@uclouvain.be}, {\tt raphael.jungers@uclouvain.be}}

\begin{abstract}
The importance of a node in a directed graph can be measured by its PageRank. The PageRank of a node is used in a number of application contexts -- including ranking websites -- and can be interpreted as the average portion of time spent at the node by an infinite random walk. We consider the problem of maximizing the PageRank of a node by selecting some of the edges from a set of edges that are under our control. By applying results from Markov decision theory, we show that an optimal solution to this problem can be found in polynomial time. 
Our core solution results in a linear programming formulation, but we also provide an alternative greedy algorithm, a variant of policy iteration, which runs in polynomial time, as well. Finally, we show that, under the slight modification for which we are given mutually exclusive pairs of edges, the problem of PageRank optimization 
becomes NP-hard.
\end{abstract}


\end{frontmatter}

% \linenumbers

\section{Introduction}
The {\em importance} of a node in a directed graph can be measured by its {\em PageRank}. The PageRank of a node \cite{Brin1998} can be interpreted as the average portion of time spent at the node by an infinite {\em random walk} \cite{Langville2006}, or in other words, the weight of the node with respect to the {\em stationary distribution} of an associated homogeneous Markov chain. PageRank is traditionally applied for ordering web-search results, but it also has many other applications \cite{Berkhin2005}, for example, in bibliometrics, ecosystems, spam detection, web-crawling, semantic networks, relational databases and natural language processing.

It is of natural interest to search for the maximum or minimum PageRank that a node (e.g., a website) can have depending on the presence or absence of some of the edges (e.g., hyperlinks) in the graph \cite{Olsen2009}. For example, since PageRank is used for ordering web-search results, a web-master could be interested in increasing the PageRank of some of his websites by suitably placing hyperlinks on his own site or by buying advertisements or making alliances with other sites \cite{Avrachenkov2006,DeKerchove2008}. Another motivation is that of estimating the PageRank of a node in the presence of {\em missing information} on the graph structure. If some of the links on the internet are broken, for example, because the server is down or there are network traffic problems, we may have only partial information on the link structure of the web-graph. However, we may still want to estimate the PageRank of a website by computing the maximum and minimum PageRank that the node may possibly have depending on the presence or absence of the unknown, hidden hyperlinks \cite{Ishii2009}. These hidden edges are often referred to as {\em fragile links}.

It is known that if we place a new edge in a directed graph, the PageRank of the terminal node of the edge can only increase. Optimal linkage strategies are known for the case in which we want to optimize the PageRank of a node and we only have access to the edges starting from this node \cite{Avrachenkov2006}. This first result has later been generalized to the case for which we are allowed to configure all of the edges starting from a given set of nodes \cite{DeKerchove2008}. 

The general problem of optimizing the PageRank of a node in the case where we are allowed to decide the absence or presence of the edges in a given {\em arbitrary} subset of edges is proposed by Ishii and Tempo \cite{Ishii2009}. They are motivated by the problem of ``fragile links'' and mention the lack of efficient, polynomial time algorithms to this problem. Then, using interval matrices, they propose an approximate solution to the problem. 

Fercoq {\em et al}.\ \cite{fercoq:ergodic} consider a continuous variant of PageRank optimization in which one can choose the intensity of the links. They allow affine coupling constraints, concave objective functions and apply convex programming.

In this paper we show that the PageRank optimization problem can be efficiently formulated as a {\em Markov decision process} (MDP), more precisely, as a {\em stochastic shortest path} (SSP) problem, and that it can therefore be solved in {\em polynomial time}. Our proof provides a {\em linear programming} formulation that can then be solved by standard techniques, but we propose a greedy algorithm, as well, which is a variant of the {\em policy iteration} algorithm.
This latter method also runs in polynomial time, under some assumptions. Our main result on polynomial-time computability remains valid even if the {\em damping constant} and the {\em personalization vector} are part of the input and it does not depend on the particular way the {\em dangling nodes} are handled. We also prove that under the slight modification for which we are given mutually exclusive constraints between pairs of edges, the problem becomes {\em NP-hard}.

%\medskip
\section{Definitions and Preliminaries} 
In this section we define the 
concept of {\em PageRank} and the PageRank optimization problem as well as give a brief introduction to stochastic shortest path problems, a special class of Markov decision processes (MDPs).

\subsection{PageRank}
Let $\mathcal{G} = \left(\mathcal{V}, \mathcal{E}\right)$ be a directed graph, where $\mathcal{V} = \left\{1, \dots, n \right\}$ is the set of vertices and $\mathcal{E} \subseteq \mathcal{V} \times \mathcal{V}$ is the set of edges. First, for simplicity, we assume that $\mathcal{G}$ is {\em strongly connected}. The adjacency matrix of $\mathcal{G}$ is denoted by $A$. Since $\mathcal{G}$ is strongly connected, $A$ is {\em irreducible}. We are going to define a {\em random walk} on the graph. If we are in node $i$, in the next step we will go to node $j$ with probability $1/deg(i)$ if $j$ is an out-neighbor of $i$, where $deg(\cdot)$ denotes out-degree. This defines a {\em Markov chain} with transition-matrix
\begin{equation}
\label{rndw}
P \triangleq \left(D_A^{-1}A\right)^\mathrm{T} \hspace{16mm}\mbox{with} \hspace{16mm} D_A \triangleq diag(A \1 )
\end{equation}
where $\1 = \left<1, \dots, 1 \right>^\mathrm{T}$ is the all-one vector and $diag(\cdot)$ is an operator that creates a diagonal matrix from a vector, more precisely, $(D_A)_{ii} \triangleq (A \1)_i = deg(i)$. Note that $P$ is a column (left) stochastic matrix and the chain can be interpreted as an infinite random walk on the graph (e.g., a random surfing).

The PageRank vector, $\prv$, of the graph is defined as the {\em stationary} {\em distribution} of the above described Markov chain, more precisely, as $P\, \prv = \prv$, where $\prv \geq 0 $ and $\prv^\mathrm{T}\1=1$. Since $P$ is an irreducible stochastic matrix, we know, e.g., from the Perron-Frobenius theorem, that $\prv$ exists and is unique.

Now, we turn to the general case, when we do not assume that $\mathcal{G}$ is strongly connected, it can be an arbitrary directed graph. In this case, there may be nodes which do not have any outgoing edges. They are usually referred to as {\em dangling nodes}. There are many ways to handle them \cite{Berkhin2005}, for example, 
we can delete them, we can add a self-loop to them, each dangling node can be linked to an artificial node (sink) or we can connect each dangling node to every other node. This last solution can be interpreted as restarting the random walk from a random starting state if we reach a dangling node. Henceforth, we will assume that we have already dealt with the dangling nodes and, therefore, every node has at least one outgoing edge.

We can then define a Markov chain similarly to (\ref{rndw}), but this chain may not have a unique stationary distribution. To solve this problem, the PageRank vector is defined as the stationary distribution of the ``Google matrix'' \cite{Langville2006} 
\begin{equation}
\label{googlem}
G \triangleq (1-c)\,P+c\,\vec{z} \1^{\!\mathrm{T}},
\end{equation}
where $\vec{z} > 0$ is a {\em personalization vector} satisfying $\vec{z}^\mathrm{T}\1=1$, and $c \in (0,1)$ is a {\em damping constant}. In practice, values between 
%$0.85$ and $0.9$ 
$0.1$ and $0.15$ 
are usually applied for $c$ and $\vec{z} = (1/n)\, \1$ \cite{Berkhin2005}.
The Markov chain defined by $G$ is {\em irreducible} and {\em aperiodic}, consequently, its stationary distribution uniquely exists and the Markov chain converges to it from any initial distribution \cite{Levin2009}.

An application of PageRank is that $\prv(i)$ can be interpreted as the ``importance'' of node $i$.
Therefore, we can use $\prv$ to define a {\em total pre-order} on the nodes of the graph by treating  $i \lesssim j$ if and only if $\prv(i) \leq \prv(j)$.

The PageRank vector can be approximated by the iteration $x_{n+1} \triangleq G \hspace{0.4mm}x_n$, where $x_0$ is an arbitrary stochastic vector, or it can be directly computed \cite{Avrachenkov2006}
\begin{equation}
\label{PR-inv}
\prv = c\,(I-(1-c)P)^{-1}\vec{z},
\end{equation}
where $I$ denotes the $n \times n$ identity matrix. Since $c \in (0,1)$ and $P$ is stochastic, matrix $I - (1-c)P$ is strictly diagonally dominant, thus invertible.

\subsection{PageRank Optimization} We will investigate a problem in which a subset of links are ``fragile'', i.e., we do not know whether they are present in the graph or we have control over them, and we want to compute the maximum (or minimum) PageRank that a specific node can have \cite{Ishii2009}. More precisely, we are given a digraph $\mathcal{G} = \left( \mathcal{V}, \mathcal{E} \right)$, a node $v \in \mathcal{V}$ and a set $\mathcal{F} \subseteq \mathcal{E}$ corresponding to those edges which are under our control. It means that we can choose which edges in $\mathcal{F}$ are present and which are absent, but the edges in $\mathcal{E} \setminus \mathcal{F}$ are fixed, they must exist in the graph. We will call any $\mathcal{F}_+ \subseteq \mathcal{F}$ a {\em configuration} of fragile links: $\mathcal{F}_+$ determines those edges that we add to the graph, while $\mathcal{F}_- = \mathcal{F} \setminus \mathcal{F}_+$ denotes those edges which we remove. The PageRank of $v$ under the $\mathcal{F}_+$ configuration is defined as the PageRank of $v$ w.r.t.\ the graph $\mathcal{G}_{\vspace{0.09mm}0} = \left( \mathcal{V}, \mathcal{E} \setminus \mathcal{F}_- \right)$. The problem is the following: how should we configure the fragile links to maximize (or minimize) the PageRank of a given node $v$?

\begin{table}[h]
\begin{center}
\hspace*{-2mm}
\footnotesize
\begin{tabular}{|l l|} \hline
\multicolumn{2}{|l|}{\textsc{The Max-PageRank Problem}\rule[4mm]{0pt}{0pt}}\\
\rule[4mm]{0pt}{0pt}{\em Instance:} & A digraph $\mathcal{G} = \left( \mathcal{V}, \mathcal{E} \right)$, a node $v \in \mathcal{V}$ and a set of controllable edges $\mathcal{F} \subseteq \mathcal{E}$.\\
{\em Optional:} & A damping constant $c \in (0,1)$ and a stochastic personalization vector $z$.\\
{\em Task:} & Compute the maximum possible PageRank of $v$ by changing the edges in $\mathcal{F}$ \\
& and provide a configuration of edges in $\mathcal{F}$ for which the maximum is taken. \vspace{-2.5mm} \\
&\\ \hline
\end{tabular}
\vspace{-5mm}
\end{center}
\label{MaxPageRank}
\end{table}

The Min-PageRank problem, which can be used, e.g., to obtain a sharp lower bound on the PageRank of a node in case the link structure is only partially known, can be stated similarly. We will concentrate on Max-PageRank, but a straightforward modification of our method can deal with the Min problem, as well. We will show that Max-PageRank can be solved in polynomial time, under the Turing model of computation, even if the damping constant and the personalization vector are part of the input, i.e., not fixed.

Of course, in particular instance of the Max-PageRank problem, there are finitely many configurations, thus, we can try to compute them one-by-one. If we have $d$ fragile links, there are $2^d$ possible graphs. The PageRank vector of a graph can be computed in $O(n^3)$ via a matrix inversion\footnote{It can be done a little faster, in $O(n^{2.376})$, using the Coppersmith-Winograd method.}.
%that we can achieve. Given a graph, we can compute 
%its PageRank vector in $O(n^3)$, via a matrix inversion (in theory, it could be done a little bit faster, in $O(n^{2.376})$, using the Coppersmith-Winograd algorithm).
The resulting 
%``brute force'' 
``exhaustive search'' algorithm has $O(\hspace{0.05mm}n^3\hspace{0.05mm} 2^d\hspace{0.05mm})$ time complexity.

%There are finitely many configurations, consequently, we can try to compute them one-by-one. If we have $d$ fragile links, then there are $2^d$ possible graphs that we can achieve. Given a graph, we can compute its PageRank vector in $O(n^3)$, via a matrix inversion (in theory, it could be done a little bit faster, in $O(n^{2.376})$, using the Coppersmith-Winograd algorithm).The resulting ``brute force'' optimization algorithm has $O(\hspace{0.05mm}n^3\hspace{0.05mm} 2^d\hspace{0.05mm})$ time complexity.
%
% \cite{Coppersmith1990}).
%\footnote{It can be done  a little bit faster, in $O(n^{2.376})$, using the Coppersmith-Winograd algorithm \citep{Coppersmith1990}.}. 
%Fortunately, as we will see in the next section, there also exist polynomial algorithms for optimizing the PageRank via the fragile links.

Note that if the graph was {\em undirected}, the Max-PageRank problem would be easy. We know \cite{Lovasz1996} that a random walk on an undirected graph, a time-reversible Markov chain, has the stationary distribution $\prv(i) = deg(i)/2m$ for all nodes $i$, where $m$ denotes the number of edges and $deg(i)$ is the degree of node $i$. Hence, 
%it is easy to see that, 
in order to maximize the PageRank of a given node $v$, we should keep edge $(i,j) \in \mathcal{F}$ if and only if $i = v$ or $j = v$.
%\vspace{-1mm}

%\medskip
\subsection{Stochastic Shortest Path Problems}
\label{MDP-complexity}
In this section we give an overview on stochastic shortest path problems,
% and their complexity
since our solutions to PageRank optimization are built upon their theory.

{\em Stochastic shortest path} (SSP) problems are generalizations of (deterministic) shortest path problems 
%\cite{Bertsekas1991,Bertsekas1996}. 
\cite{Bertsekas1996}. 
In an SSP problem the transitions between the nodes are uncertain, but we have some control over their probability distributions. 
%We aim at finding a control policy (a function from nodes to controls) such that by applying this policy we can reach a given target state with probability one while minimizing the expected costs, as well. 
We aim at finding a control policy (a function from nodes to controls) that minimizes the expected (cumulative) cost of reaching a given target state.
SSP problems are 
%special 
finite, undiscounted {\em Markov decision processes} (MDPs) with 
%finite state and action spaces and with 
an absorbing, cost-free termination state. 
%They are of high practical importance since they arise in many real-world domains, for example, several problems in operations research can be formulated as an SSP problem or contains it as a sub-problem. Typical examples are routing and resource allocation problems.

%\subsubsection{Problem Formulation}
An SSP problem can be stated as follows. We have given a finite set of {\em states}, $\mathbb{S}$, and a finite set of control {\em actions}, $\mathbb{U}$. For simplicity, we assume that $\mathbb{S} = \left\{1,\dots, n, n+1\right\}$, where $\tau = n+1$ is a special state, the {\em target} or {\em termination} state. In each state $i$ we can choose an action $u \in \mathcal{U}(i)$, where $\mathcal{U}(i) \subseteq \mathbb{U}$ is the set of allowed actions in state $i$. After the action was chosen, the system moves to state $j$ with probability $p(j\,|\, i, u)$ and we incur cost $g(i, u, j)$. The cost function is real valued and the transition-probabilities are, of course, nonnegative as well as they sum to one for each state $i$ and action $u$. 
%Namely, for all state $i$, $j$ and action $u \in \mathcal{U}(i)$,
%% we have that
%$p(j \,|\, i, u) \geq 0$ and
%\begin{equation}
%\sum_{j=1}^{n+1}p(j \,|\, i, u) = 1.
%\end{equation}
The target state is {\em absorbing} and {\em cost-free} that is, if we reach state $\tau$, we remain there forever without incurring any more costs. More precisely, for all 
%action 
$u \in \mathcal{U}(\tau)$, 
%we have 
$p(\tau \,|\, \tau, u) = 1$ and $g(\tau,u,\tau) = 0$.
%\begin{equation}
%\label{absorbing}
%p(\tau \,|\, \tau, u) = 1, \hspace{16mm} g(\tau,u,\tau) = 0.
%\end{equation}

The problem is to find a control {\em policy} such that it reaches state $\tau$ with probability one and minimizes the expected 
%cumulative 
costs, as well. A (stationary, Markov) {\em deterministic} policy is a function from states to actions, $\mu: \mathbb{S} \to \mathbb{U}$. A {\em randomized} policy can be formulated as $\mu: \mathbb{S} \to \Delta(\mathbb{U})$, where $\Delta(\mathbb{U})$ denotes the set of all probability distributions over set $\mathbb{U}$. It can be shown that every such policy induces a {\em Markov chain} on the state space \cite{MDP2002}. A policy is called {\em proper} if, using this policy, the termination state will be reached with probability one, and it is {\em improper} otherwise. The {\em value} or {\em cost-to-go} function of policy $\mu$ gives us the expected total costs 
%that we incur 
of starting from a state and following $\mu$ thereafter; that is, 
%it is defined as
\begin{equation}
J^{\mu}(i) \triangleq \lim_{k \to \infty} \mathbb{E}_{\hspace{0.3mm}\mu}\! \left[\,\sum_{t=0}^{k-1} g(i_t, u_t, i_{t+1}) \biggm| i_0 = i \,\right],
\end{equation}
for all states $i$, where $i_t$ and $u_t$ are random variables representing the state and the action taken at time $t$, respectively. Naturally, 
%state 
$i_{t+1}$ is of distribution $p(\cdot\,|\, i_t, u_t)$ and 
%action 
$u_t$ is of distribution $\mu(i_t)$; or $u_t = \mu(i_t)$ in case we apply a deterministic policy.
Note that by applying a proper policy, we arrive at a finite horizon problem, however, the length of the horizon may be random and may depend on the applied control policy, as well.

We say that $\mu_1 \leq \mu_2$ if and only if for all states $i$, $J^{\mu_1}(i) \leq J^{\mu_2}(i)$. A policy is (uniformly) {\em optimal} if it is better than or equal to all other policies. There may be many optimal policies, but assuming that (A1) there exists at least one proper policy and (A2) every improper policy yields infinite cost for at least one initial state, they all share the same unique optimal value function, $J^*$. Then, function $J^*$ is the unique solution of the {\em Bellman optimality equation}, $TJ^* = J^*$, where $T$ is the {\em Bellman operator} \cite{Bertsekas1996}, that is,
\begin{equation}
\label{Bellman-opt-equation}
(TJ)(i) \triangleq \min_{u \in \mathcal{U}(i)}\,{\sum_{j = 1 }^{n+1}{\,p(j\,|\,i,u) \Bigl[\,g(i,u,j )+ J(j) }\,\Bigr]},
\end{equation}
for all states $i \in \mathbb{S}$ and value functions $J: \mathbb{S} \to \mathbb{R}$.
The Bellman operator of a (randomized) policy $\mu$ is defined for all state $i$ as
\begin{equation}
\label{Bellman-opt-equation}
(T_{\mu}J)(i) \triangleq \!\! \sum_{u \in \mathcal{U}(i)}\!\! \mu(i, u)\,{\sum_{j = 1 }^{n+1}{\,p(j\,|\,i,u)\Bigl[\,g(i,u,j)+ J(j)}\,\Bigr]},
\end{equation}
where $\mu(i, u)$ is the probability that policy $\mu$ chooses action $u$ in state $i$. 

Given the assumptions above, {\em value iteration} converges in SSPs \cite{Bertsekas1991}, 
\begin{equation}
\lim_{k \to \infty}{T_{\mu}^k J} = J^{\mu},  \hspace{16mm} \lim_{k \to \infty}{T^k J} = J^{*}.
\end{equation}

Operators $T$ and $T_{\mu}$ are {\em monotone} and, assuming that (APP) all policies are proper,
$T$ and $T_{\mu}$ are {\em contractions} w.r.t.\ a weighted maximum norm \cite{Bertsekas1996}.

From a given value function $J$, it is straightforward to get a policy, e.g., by applying a {\em greedy} policy \cite{Bertsekas1996} with respect to $J$ that is, for all state $i$,
\begin{equation}
\mu(i) \in \argmin_{u \in \mathcal{U}(i)}\,{\sum_{j = 1}^{n+1}{\,p(j\,|\,i,u)\Bigl[\,g(i, u, j) + J(j)}\,\Bigr]}.
\end{equation}

There are several solution methods for solving MDPs, e.g., in the fields of {\em reinforcement learning} and [neuro-] {\em dynamic programming}.
Many of these algorithms aim at finding (or approximating) the optimal value function, since good approximations to $J^*$ directly lead to good policies \cite{Bertsekas1996}. General solution methods include value iteration, policy iteration, Gauss-Seidel method, Q-learning, SARSA and TD($\lambda$):\ temporal difference learning \cite{Bertsekas1996,MDP2002,Sutton1998}.

Later, we will apply a variant of the {\em policy iteration} (PI) algorithm. The basic version of PI works as follows. We start with an arbitrary {\em proper} policy, $\mu_0$. In iteration $k$ we first {\em evaluate} the actual policy, $\mu_k$, by solving the linear system, $T_{\mu_k}J^{\mu_k} = J^{\mu_k}$, and then we {\em improve} the policy by defining $\mu_{k+1}$ as the greedy policy w.r.t.\ $J^{\mu_k}$. The algorithm terminates if $J^{\mu_k} = J^{\mu_{k+1}}$. Assuming (A1) and (A2), PI generates an improving sequence of proper policies and it always finds an optimal solution in a finite number of iterations \cite{Bertsekas1996}.

It is known that all of the three classical variants of MDPs (finite horizon, infinite horizon discounted cost and infinite horizon average cost) can be solved in polynomial time \cite{Littman95}. Moreover, these classes of problems are P-complete \cite{Papadimitriou1987}. 
In the case of SSP problems, they can be reformulated as {\em linear programming} (LP) problems \cite{Bertsekas1996}, more precisely, the optimal cost-to-go,
$J^*(1), \dots, J^*(n)$, solves the following LP in variables $x_1, \dots, x_n$\,:
\begin{subequations}
\begin{align}
\mbox{maximize}& \hspace{10mm}
\sum_{i=1}^{n}\,x_i\\
\mbox{subject to}& \hspace{10mm}
x_i \, \leq \, \sum_{j = 1 }^{n+1}{\,p(j\,|\,i,u) \Big[\,g(i,u,j) + x_j\Big]}
\end{align}
\end{subequations}
for all states $i$ and actions $u \in \mathcal{U}(i)$. Note that the value of the termination state, $x_{n+1}$, is fixed at zero. This LP has $n$ variables and $O(n m)$ constraints, where $m$ is the maximum number of allowed actions per state. Knowing that an LP can be solved in polynomial time \cite{Gonzaga1988} (in the number of variables, the number of constraints and the binary size of the input), this reformulation already provides a way to solve an SSP problem in {\em polynomial time}.

Assuming that all policies are proper (APP), the state space can be {\em partitioned} into nonempty subsets $S_1, \dots, S_r$ such that for any $1 \leq q \leq r$, state $i \in S_q$ and action $u \in \mathcal{U}(i)$, there exists some $j \in \{\tau\}\cup S_1 \cup \dots \cup S_{q-1}$ such that $p(j\,|\,i,u)>0$. Then, if assumption (APP) holds, value iteration can find an optimal policy after a number of iterations that is bounded by a polynomial in $L$ (the binary input size) and $\eta^{-2r}$, where $\eta$ is the smallest positive transition probability \cite{Tseng1990}. 
Since policy iteration converges no more slowly than value iteration \cite{Puterman1994}, policy iteration also terminates in iterations bounded by a polynomial in $L$ and $\eta^{-2r}$, assuming (APP).

%\medskip
\section{PageRank Optimization as a Markov Decision Process}
Before we prove that efficient algorithms to Max-PageRank do exist, first, we recall a basic fact about stationary distributions of Markov chains.

Let $(X_0, X_1, \dots)$ denote a time-homogeneous Markov chain defined on a finite set $\Omega$. The {\em expected first return time} of a state $i \in \Omega$ is defined as 
\begin{equation}
\frt(i) \triangleq \mathbb{E} \left[ \, \inf  \left\{\, t \geq 1 : X_t = i\, \right\} \,|\, X_0 = i \,\right].
\end{equation}
If state $i$ is {\em recurrent}, then $\varphi(i)$ is finite. Moreover, if the chain is irreducible,
\begin{equation}
\label{apt}
\prv(i) = \frac{1}{\frt(i)},
\end{equation}
for all states $i$, where $\prv$ is the stationary distribution of the Markov chain \cite{Levin2009}. This naturally generalizes to {\em unichain} processes, viz., when we have a single {\em communicating class} of states and possibly some {\em transient} states. In this case we need the convention that $1/\infty = 0$, since the expected first return time to transient states is $\infty$. Hence, the stationary distribution of state $i$ can be interpreted as the {\em average portion of time} spent in $i$ during an infinite random walk. It follows from 
equation (\ref{apt}) that the problem of 	maximizing [minimizing] the PageRank of a node is equivalent to the problem of minimizing [maximizing] the expected first return time to this node.

We will show that the Max-PageRank problem can be efficiently formulated as a {\em stochastic shortest path} (SSP) problem \cite{Bertsekas1996}, where ``efficiently'' means that the construction (reduction) takes polynomial time. First, we will consider the PageRank optimization {\em without damping}, namely $c = 0$, but later, we will extend the model to the case of damping and personalization, as well. We will start with a simple, but intuitive reformulation of the problem. Though, this reformulation will not ensure that Max-PageRank can be solved in polynomial time, it is good to demonstrate the main ideas and to motivate the refined solution.

\subsection{Assumptions}
\label{assumptions}
First, we will make two assumptions, in order to simplify the presentation of the construction, but later, in the main theorem, they will be relaxed.
%\vspace{1mm}
\begin{enumerate}
\item[(AD)] {\em Dangling Nodes Assumption}\,: We assume that there is a fixed (not fragile) outgoing edge from each node of the graph. This assumption guarantees that there are no dangling nodes as well as there are no nodes with only fragile links (which would be latent dangling nodes).
%\vspace{2mm}
\item[(AR)] {\em Reachability Assumption}\,: We also assume that for at least one configuration of fragile links we have a unichain process and node $v$ is recurrent, namely, we can reach node $v$ with positive probability from all nodes of the graph. This assumption is required to have a well-defined PageRank for at least one configuration. In our SSP formulation this 
assumption will be equivalent to assuming that there is at least one {\em proper} policy.
In case of damping, this assumption is automatically satisfied, as the Markov chain will be
irreducible, and accordingly, unichain. On the other hand, irrespective of how we
configure fragile links, all policies in the corresponding SSP problem are proper.
\end{enumerate}

\subsection{Simple SSP Formulation}
\label{simplef}
First, let us consider an instance of Max-PageRank. We are going to build an associated SSP problem that solves the original PageRank optimization problem. The {\em states} of the MDP are the nodes of the graph, except for $v$ which we ``split'' into two parts and replace by two new states: $v_s$ and $v_t$. Intuitively, state $v_s$ will be our ``starting'' state: it has all the outgoing edges of $v$ (both fixed and fragile), but it does not have any incoming edges. The ``target'' state will be $v_t$: it has all the incoming edges of node $v$ and, additionally, it has only one outgoing edge: a self-loop. Note that $\tau = v_t$, namely, $v_t$ is the {\em absorbing termination state} of the associated SSP problem.

\vspace{3mm}
\begin{figure}[h]
\begin{center}
\resizebox{10cm}{!}{\includegraphics*{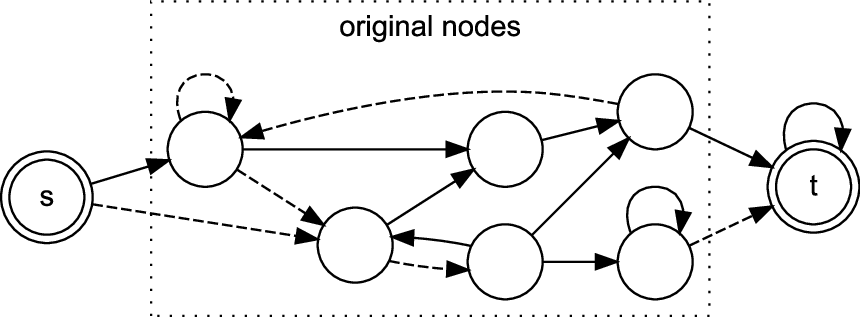}}
\vspace{1mm}
\caption{SSP reformulation: the starting state is $s = v_s$, the target state is $t = v_t$ and the dashed edges denote fragile links. The original nodes in the rectangle exclude $v$.}
\end{center}
\vspace{-2mm}
\end{figure}

An {\em action} in state $i$ is to select a subset of fragile links (starting from $i$) which we ``turn on'' (activate). All other fragile links from $i$ will be ``turned off'' (deactivated). Thus, in state $i$ the allowed set of actions is $\mathcal{U}(i) \triangleq \mathcal{P}(\mathcal{F}_i)$, where $\mathcal{P}$ is the power set and $\mathcal{F}_i$ is the set of outgoing fragile links from $i$.

Let us assume that we are in state $i$, where there are $a_i \geq 1$ fixed outgoing edges and we have activated $b_i(u) \geq 0$ fragile links, determined by action $u \in \mathcal{U}(i)$. Then, the {\em transition-probability} to all states $j$ that can be reached from state $i$ using a fixed or an activated fragile link is $p(j \,|\, i, u) \triangleq 1/(a_i+b_i(u))$.

We define the {\em immediate-cost} of all actions as one, except for the actions taken at the cost-free target state. Thus, the immediate-cost function is
\begin{equation}
g(i, u, j) \triangleq
\left\{
\begin{array}{ll}
0 \hspace{3mm}& \mbox{if}\hspace{2mm} i = v_t,\\
1 \hspace{3mm}& \mbox{otherwise},\\
\end{array}
\right.
\end{equation}
for all states $i$, $j$ and actions $u$. Note that taking an action can be interpreted as performing a step in the random walk. Therefore, the expected cumulative cost of starting from state $v_s$ until we reach the target state $v_t$ is equal to the expected number of steps until we first return to node $v$ according to our original random walk. It follows, that the above defined SSP formalizes the problem of {\em minimizing} 
%(via a configuration) 
the expected first return time to state $v$. 
Hence, its solution is equivalent to {\em maximizing} the PageRank of node $v$.

Each allowed deterministic {\em policy} $\mu$ defines a potential way to configure the fragile links. Moreover, the $v_s$ component of the cost-to-go function, $J^{\mu}(v_s)$, is the expected first return time to $v$ using the fragile link configuration of $\mu$. Therefore, we can compute the PageRank of node $v$ by 
\begin{equation}
\prv(v) = \frac{1}{J^{\mu}(v_s)},
\end{equation}
where we applied the convention of $1/\infty = 0$, which is needed when $v$ is not recurrent under $\mu$. Thus, the maximal PageRank of $v$ is $1/J^*(v_s)$.

Most solution algorithms compute the optimal cost-to-go function, $J^*$, but even if we use a direct policy search method, it is still easy to get back the value function of the policy. We can compute, for example, the expected first return time if we configure the fragile links according to policy $\mu$ as follows. For simplicity, assume that $v_s = 1$, then
\begin{equation}
\label{j1}
J^{\mu}(1) = \1^{\!\mathrm{T}}(I - P_{\mu})^{-1}e_1,
\end{equation}
where $e_j$ is $j$-th canonical basis vector, $I$ is an $n \times n$ identity matrix and $P_{\mu}$ is the {\em substochastic} transition matrix of the corresponding SSP problem {\em without the row and column of the target state}, $v_t$, if we configure the fragile links according to policy $\mu$.
Regarding the invertibility of $I - P_{\mu}$ note that
\begin{equation}
(I - P_{\mu})^{-1} = \sum_{n=0}^{\infty}P_{\mu}^{\hspace{0.3mm}n},
\end{equation}
and we know that this {\em Neumann series} converges if $\varrho(P_{\mu}) < 1$, where $\varrho(\cdot)$ denotes {\em spectral radius}. Thus, $(I - P_{\mu})^{-1}$ is well-defined for all {\em proper} policies, since it is easy to see that policy $\mu$ is proper if and only if $\varrho(P_{\mu}) < 1$.

It is known that MDPs can be solved in polynomial time in the number of states, $N$, and the maximum number of actions per state, $M$ (and the maximum number of bits required to represent the components, $L$), e.g., by linear programming \cite{Littman95,Papadimitriou1987}. The size of the state space of the current formulation is $N = n+1$, where $n$ is the number of vertices of the original graph, but, unfortunately, its action space does not have a polynomial size. For example, if we have maximum $m$ fragile links leaving a node, we have $2^m$ possible actions to take, namely, we could switch each fragile link independently on or off, consequently, $M = 2^m$. Since $m = O(n)$, from the current reformulation of problem, we have that there is a solution which is polynomial but in $2^n$, which is obviously not good enough. However, we can notice that if we restrict the maximum number of fragile links per node to a constant, $k$, then we could have a solution which is polynomial in $n$ (since the maximum number of actions per state becomes constant: $2^k$). This motivates our refined solution, in which we reduce the maximum number of actions per state to two while only slightly increasing the number of states.

\subsection{Refined SSP Formulation}
\label{subsect-refined-SSP}
Now, we present a refined SSP formulation which will be the base of the proof that shows the polynomial time computability of Max-PageRank.

We are going to modify our previous SSP formulation. The key idea will be to introduce an {\em auxiliary state} for each fragile link. Therefore, if we have a fragile link from node $i$ to node $j$ in the original graph, we place an artificial state, $f_{ij}$, ``between'' them in the refined reformulation. The refined transition-probabilities are as follows. Let us assume that in node $i$ there were $a_i \geq 1$ fixed outgoing edges and $b_i \geq 0$ fragile links. Now, in the refined formulation, in state $i$ we have only one available action which brings us uniformly, with $1/(a_i+b_i)$ probability, to state $j$ or to state $f_{ij}$ depending respectively on whether there was a fixed or a fragile link between $i$ and $j$. Notice that this probability is {\em independent} of how many fragile links are turned on, it is always the same. In each auxiliary state $f_{ij}$ we have two possible actions: we could either turn the fragile link on or off. If our action is ``on'' (activation), we go with {\em probability one} to state $j$, however, if our action is ``off'' (deactivation), we return with {\em probability one} to state $i$.

\begin{figure}[h]
\begin{center}
%\resizebox{10cm}{!}{\includegraphics*{ssp-trick1.eps}}
\resizebox{9.5cm}{!}{\includegraphics*{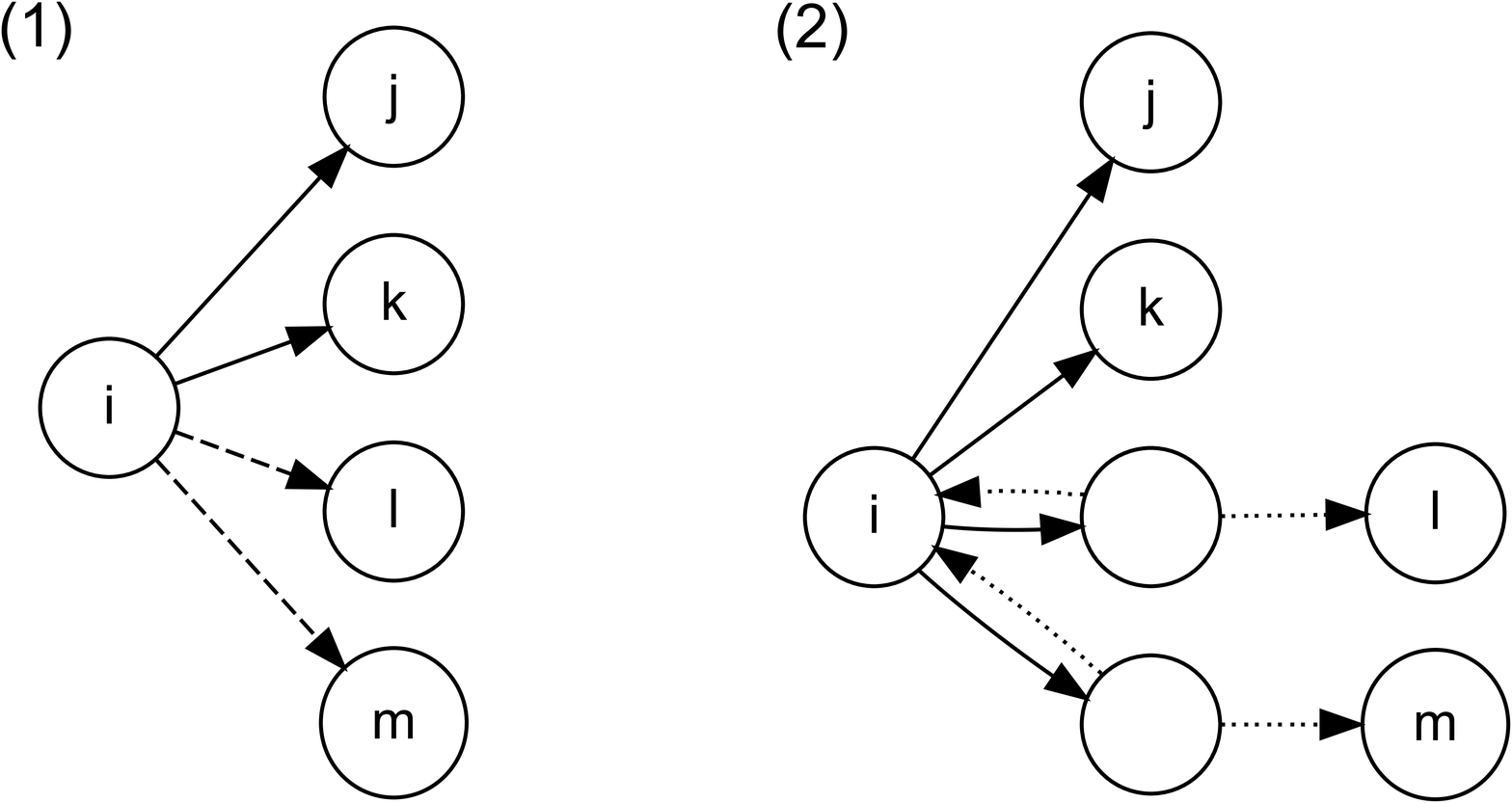}}
\vspace{2mm}
\caption{An example for inserting auxiliary states for fragile links. The left hand side presents the original situation, in which dashed edges are fragile links. The right hand side shows the refined reformulation, where the dotted edges represent possible actions.}
\label{f2}
\vspace{-3mm}
\end{center}
\end{figure}

We should check whether the transition-probabilities between the original nodes of graph are not affected by this reformulation (it is illustrated by Figure \ref{f2}). Suppose, we are in node $i$, where there are $a$ fixed and $b$ fragile links\footnote{For simplicity, now we do not denote their dependence on node $i$.}, and we have turned $k$ of the fragile links on. Then, the transition-probability to each node $j$, which can be reached via a fixed or an activated fragile link, should be $1 / (a + k)$. In our refined reformulation, the immediate transition-probability from state $i$ to state $j$ is $1/(a + b)$, however, we should not forget about those $b-k$ auxiliary nodes in which the fragile links are deactivated and which lead back to state $i$ with probability one, since, after we returned to state $i$ we have again $1/(a + b)$ probability to go to state $j$ an so on. Now, we will compute the probability of eventually arriving at $j$ if we start in $i$ and only visit auxiliary states meantime.

\vspace{2mm}

To simplify the calculations, let us temporarily replace each edge leading to an auxiliary state corresponding to a {\em deactivated} fragile link with a self-loop. We can safely do so, since these states lead back to state $i$ with probability one, therefore, the probability of eventually arriving at node $j$ does not change by this modification. After this modification, the probability of arriving at state $j$ if one starts in state $i$ can be written as
\begin{subequations}
\begin{equation}
\mathbb{P}\left(\,\exists \, t: X_t = j\,|\, \forall \,s < t: X_s = i\,\right)\, = 
\vspace{2mm}
\end{equation}
\begin{equation}
= \,\sum_{t=1}^{\infty}\,\mathbb{P}\left(\, X_t = j\,|\, X_{t-1} = i \, \right)\, \prod_{s=1}^{t-1}\mathbb{P}\left(\,  X_s = i \,|\, X_{s-1}=i\,\right) \, = 
\vspace{2mm}
\end{equation}
\begin{equation}
= \,\sum_{t=1}^{\infty}\frac{1}{a+b} \left(\frac{b-k}{a+b}\right)^{\!t-1}\! = \,\frac{1}{a+b}\,\sum_{t=0}^{\infty} \left(\frac{b-k}{a+b}\right)^{\!t} = \,	\frac{1}{a+k}\,.
\end{equation}
\end{subequations}
\vspace{1mm}

With this, we proved that the probability of eventually arriving at state $j$ if we start in state $i$, before arriving at any (non-auxiliary) state $l$ that was reachable via a fixed or a fragile link from $i$ in the original graph, is the same as the one-step transition-probability was from state $i$ to state $j$ according to the original random walk. This partially justifies the construction.

However, we should be careful, since we might have performed several steps in the auxiliary nodes before we finally arrived at state $j$. Fortunately, this phenomenon does not ruin our ability to optimize the expected first return time to state $v$ in the original graph, since we count the steps with the help of the cost function, which can be refined according to our needs. All we have to do is to allocate zero cost to those actions which lead us to auxiliary states.
More precisely, the immediate-cost function should be
\begin{equation}
\label{ref-cost}
g(i, u, j) \triangleq
\left\{
\begin{array}{ll}
0 \hspace{3mm}& \mbox{if}\hspace{2mm} i = v_t \hspace{2mm}\mbox{or}\hspace{2mm} j = f_{il} \hspace{2mm}{or}\hspace{2mm}u=\mbox{``off''},\\
1 \hspace{3mm}& \mbox{otherwise},\\
\end{array}
\right.
\end{equation}
for all states $i$, $j$, $l$ and action $u$. Consequently, we only incur cost if we directly go from state $i$ to state $j$, without visiting an auxiliary node (it was a fixed link), or if we go to state $j$ via an activated fragile link, since we have $g(f_{ij},u,j)=1$ if $u=\mbox{``on''}$. It is easy to see that in this way we only count the steps of the original random walk and, e.g., it does not matter how many times we visit auxiliary nodes, since these visits do not have any cost.

This reformulation also has the nice property that $J^{\mu}(v_s)$ is the expected first return time to node $v$ in the original random walk, in case we have configured the fragile links according to policy $\mu$. The minimum expected first return time that can be achieved with suitably setting the fragile links is $J^*(v_s)$, where $J^*$ is the optimal cost-to-go function of the above constructed SSP problem. Thus, the {\em maximum} PageRank node $v$ can have is $1/J^*(v_s)$. 

It is also easy to see that if we want to compute the {\em minimum} possible PageRank of $v$, we should simply define a new immediate-cost function as $\hat{g}=-\,g$, where $g$ is defined by equation (\ref{ref-cost}). If the optimal cost-to-go function of this modified SSP problem is $\hat{J}^*$, the {\em minimum} PageRank $v$ can have is $1/|\hat{J}^*(v_s)|$. Thus, Min-PageRank can be handled with the same construction.

The number of states of this formulation is $N=n+d+1$, where $n$ is the number of nodes of the original graph and $d$ is the number of fragile links. Moreover, the maximum number of allowed actions per state is $M = 2$, therefore, this SSP formulation provides a proof that, assuming (AD) and (AR), Max-PageRank can be solved in {\em polynomial time}. The resulted SSP problem can be reformulated as a linear program, namely, the optimal cost-to-go function solves the following LP in variables $x_i$ and $x_{ij}$,
\begin{subequations}
\begin{align}
\mbox{maximize}& \hspace{10mm}
\sum_{i \in \mathcal{V}}\,x_i\,+\!\sum_{(i,j)\in\mathcal{F}}{\!\!x_{ij}}\\
\mbox{\rule[5mm]{0mm}{0mm}subject to}& \hspace{10mm} x_{ij} \leq x_i\,, \hspace{6mm}\mbox{and} \hspace{6mm}x_{ij} \leq x_j+1\,,\hspace{6mm}\mbox{and}\\
\mbox{\rule[9mm]{0mm}{0mm}}& \hspace{10mm} x_i \, \leq \, \frac{1}{deg(i)} \Bigg[\,\, \sum_{(i,j) \in \mathcal{E}\setminus \mathcal{F}}{\!\!\!\!(x_j+1)}\,\, +\! \sum_{(i,j) \in \mathcal{F}}{\!\!x_{ij}}\,\, \Bigg],
\end{align}\label{linf}
\end{subequations}

\noindent for all $i \in \mathcal{V} \setminus \{ v_t \}$ and $(i,j) \in \mathcal{F}$, where $x_i$ is the cost-to-go of state $i$, $x_{ij}$ relates to the auxiliary states of the fragile edges, and $deg(\cdot)$ denotes out-degree including both fixed and fragile links (independently of the configuration). Note that we can only apply this LP after state $v$ was ``splitted'' into a starting and a target state and the value of the target state, $x_{v_t}$, is fixed at zero, since it is the termination state of the constructed SSP problem.

\subsection{Handling Dangling Nodes} 
\label{dangling}
Now, we will show that assumption (AD) can be omitted and our complexity result is independent of how dangling nodes are particularly handled.

Suppose that we have chosen a rule according to which the dangling nodes are handled, e.g., we take one of the rules discussed by Berkhin \cite{Berkhin2005}. Then, in case (AD) is not satisfied, we can simply apply this rule to the dangling nodes before the optimization. However, we may still have problems with the nodes which only have fragile links, since they are latent dangling nodes, namely, they become dangling nodes if we deactivate all of their outgoing edges. We call them ``fragile nodes''. Notice that in each fragile node we can safely restrict the optimization in a way that maximum one of the fragile links can be activated. This does not affect the optimal PageRank of $v$, since the only link allowed should point to a node that has the smallest expected hitting time to $v$. Even if there are several nodes with the same value, we can select one of them arbitrarily. Naturally, this restriction of the optimization to only one allowed activated fragile link per state is only suitable for fragile nodes, it is not applicable in general, when the node has fixed edges, as well.

It may also be the case that deactivating all of the edges is the optimal solution, for example, if the fragile links lead to nodes that have very large hitting times to $v$. In this case, we should have an action that has the same effect as the dangling node handling rule. Consequently, in case we have a fragile node that has $m$ fragile links, we will have $m+1$ available actions: $u_1, \dots, u_{m+1}$. If $u_j$ is selected, where $1 \leq j \leq m$, it means that only the $j$-th fragile link is activated and all other links are deactivated, while if $u_{m+1}$ is selected, it means that all of the fragile links are deactivated and auxiliary links are introduced according to the selected dangling node handling rule.
If we treat the fragile nodes this way, we still arrive at an MDP which has states and actions polynomial in $n$ and $d$, therefore, the PageRank optimization problem can be solved in polynomial time even if assumption (AD) is not satisfied and independently of the applied rule. The modification of the LP formulation if fragile nodes are allowed is straightforward.

\subsection{Damping and Personalization}
\label{personalization}
Now, we will extend our refined SSP formulation, in order to handle {\em damping}, as well. For the sake of simplicity, we will assume (AD), but it is easy to remove it in a similar way as it was presented in Section \ref{dangling}. Note that assumption (AR) is always satisfied in case of damping (cf.\ Section \ref{assumptions}).

Damping can be interpreted as in each step we continue the random walk with probability $1-c$ and we restart it (``zapping'') with probability $c$, where $c \in (0,1)$ is a given damping constant. In this latter case, we choose the new starting state according to the probability distribution of a given positive and stochastic personalization vector $\vec{z}$. In order to model this, we introduce a new global auxiliary state, $q$, which we will call the {\em teleportation state}, since random restarting is sometimes referred to as ``teleportation'' \cite{Langville2006}.

\begin{figure}[h]
\begin{center}
\resizebox{\columnwidth}{!}{\includegraphics*{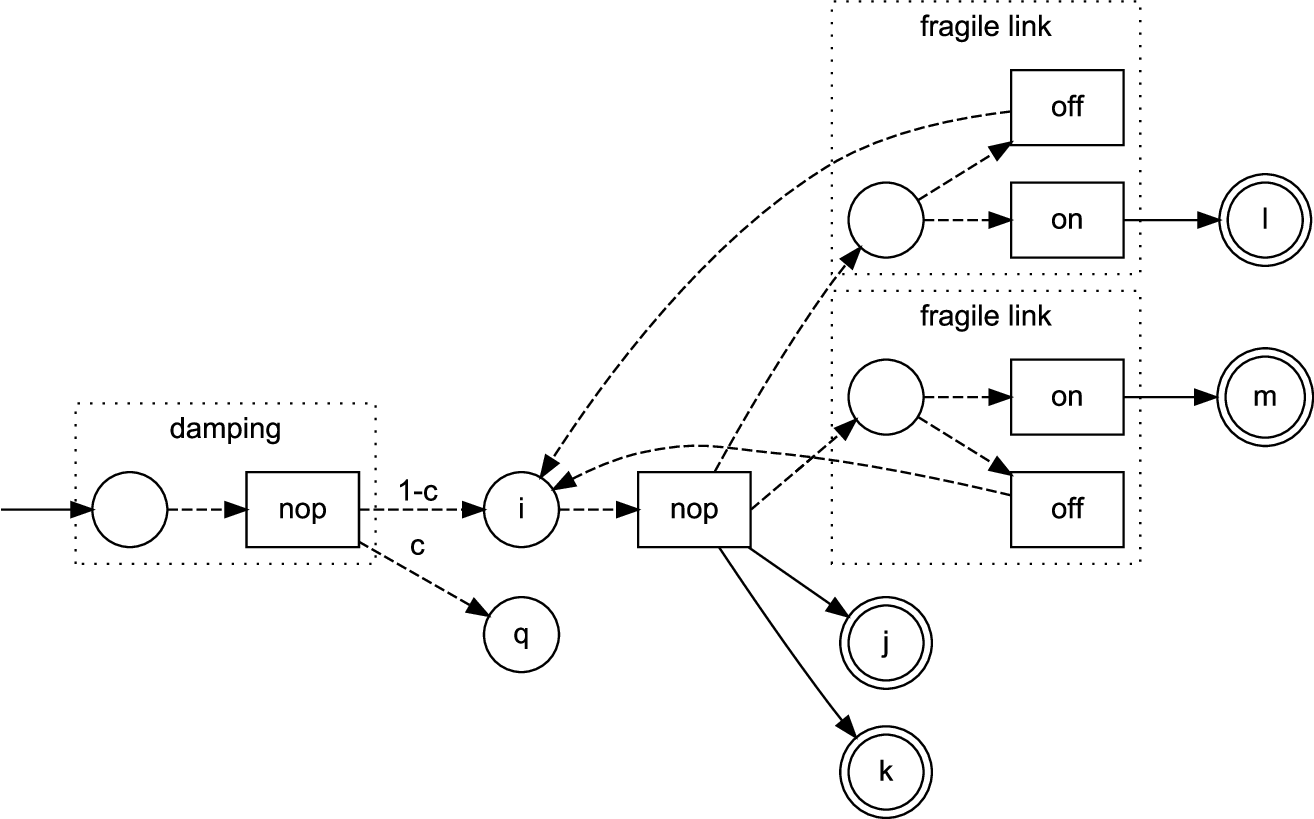}}
\caption{An illustration of damping: the substructure of a node of the original digraph. Circles represent states and boxes represent actions. State $q$ denotes the global ``teleportation'' state. Dashed edges help determining zero cost events: if a state-action-state path has only dashed edges, then this triple has zero cost, otherwise, its cost is one.}
\label{f3}
\vspace{-1mm}
\end{center}
\end{figure}

In order to take the effect of damping into account in each step, we place a new auxiliary state $h_i$ ``before'' each (non-auxiliary) state $i$ (see Figure \ref{f3}). Each action that leads to $i$ in the previous formulation now leads to $h_i$. In $h_i$ we have only one available action (``nop'' abbreviating ``no operation'') which brings us to node $i$ with probability $1-c$ and to the teleportation state $q$ with probability $c$, except for the target state, $v_t$, for which $h_{v_t}$ leads with probability one to $v_t$. In state $q$, we have only one available action which brings us with distribution $\vec{z}$ to the newly defined nodes, that is we have
\begin{equation}
p(\,h_i\,|\, q\,) \triangleq p(\,h_i\,|\, q, u\,) \triangleq
\left\{
\begin{array}{ll}
\vec{z}(i) \hspace{3mm}& \mbox{if}\hspace{2mm} i \neq v_s \hspace{2mm}\mbox{and}\hspace{2mm} i \neq v_t\\
\vec{z}(v) \hspace{3mm}& \mbox{if}\hspace{2mm} i = v_t \\
0 \hspace{3mm}& \mbox{if}\hspace{2mm}i = v_s.\\
\end{array}
\right.
\end{equation}
All other transition-probabilities from $q$ are zero. Regarding the cost function: it is easy to see that we should not count the steps when we move through $h_i$, therefore, $g(h_i,u,i)=0$ and $g(h_i,u,q)=0$. However, we should count when we move out from state $q$, i.e., $g(q, u, i) = 1$ for all $i$ and $u$.

\vspace{4mm}

The straightforward solution, namely, to connect $i$ directly to $q$ without an additional auxiliary state, $h_i$, does not work, since the transition-probability to $q$ should be constant (i.e., equal to $c$), but the probabilities of taking a link starting from $i$ change as we change the configuration of fragile links.

In this variant, in which we take damping and personalization into account,
the size of the state space is $N = 2n+d+2$ and we still have maximum $2$ actions per state, therefore, it can also be solved in {\em polynomial time}.

In this case, the LP formulation of finding the optimal cost-to-go is
\begin{subequations}
\begin{align}
\vspace{1mm}
\mbox{maximize}& \hspace{10mm}
\sum_{i \in \mathcal{V}}\,(x_i+\hat{x}_i)\,+\!\sum_{(i,j)\in\mathcal{F}}{\!\!x_{ij}}+\,x_q\\
\mbox{\rule[5mm]{0mm}{0mm}subject to}& \hspace{10mm} x_{ij} \leq \hat{x}_j+1\,,\hspace{6mm}\mbox{and} \hspace{6mm} \hat{x}_i \leq (1-c)\, x_i + c \,x_q\,,\\
\mbox{\rule[9mm]{0mm}{0mm}}& \hspace{10mm} x_{ij} \leq x_i\,,\hspace{12.2mm}\mbox{and} \hspace{6mm} x_q \, \leq \, \sum_{i \in \mathcal{V}}{\,\hat{z}_i\,(\hat{x}_i+1)}\,,\\
\mbox{\rule[6mm]{0mm}{0mm}}& \hspace{10mm} x_i \, \leq \, \frac{1}{deg(i)} \Bigg[\,\, \sum_{(i,j) \in \mathcal{E}\setminus \mathcal{F}}{\!\!\!\!(\hat{x}_j+1)}\,\, +\! \sum_{(i,j) \in \mathcal{F}}{\!\!x_{ij}}\,\, \Bigg],
\vspace{1mm}
\end{align}
\end{subequations}
for all $i \in \mathcal{V}\setminus \{ v_t\}$ and $(i,j) \in \mathcal{F}$, where $\hat{z}_i = p(\,h_i\,|\,q)$, $\hat{x}_i$ denotes the cost-to-go of $h_i$ and $x_q$ is the value of the teleportation state. All other notations are the same as in (\ref{linf}) and we also have that $x_{v_t}$ and $\hat{x}_{v_t}$ are fixed at zero.

The above LP problem has $O(n+d)$ variables and $O(n+d)$ constraints, which can thus be solved in $O((n+d)^3L)$, where $L$ is the binary input size (for rational coefficients) or in $O((n+d)^3\log \frac{1}{\varepsilon})$, where $\varepsilon$ is the desired precision \cite{Gonzaga1988}. The result that was proved through Sections \ref{subsect-refined-SSP}, \ref{dangling} and \ref{personalization} is

\begin{theorem} The \textsc{Max-PageRank Problem} can be solved in polynomial time under the Turing model of computation even if the damping constant and the personalization vector are part of the input.
\end{theorem}

Assumptions (AD) and (AR) are not needed for this theorem, since dangling and fragile nodes can be treated as discussed in Section \ref{dangling} (without increasing the complexity) and, in case of damping, all policies are proper.

Assuming that $c$ and $\vec{z}$ can be represented using a number of bits polynomial in $n$, which is the case in practice, since $c$ is usually $0.1$ or $0.15$ and $\vec{z} = (1/n)\, \1$ \cite{Berkhin2005}, we arrive at a {\em strongly} polynomial time solution, because all other coefficients can be represented using $O(\log n)$ bits. 

%\medskip
\subsection{State Space Reduction}
In practical situations the state space may be very large which can make direct solutions impractical. Approximate, sampling based methods are usually preferred in these circumstances \cite{Bertsekas1996,MDP2002,Sutton1998}. In this section, we show that the state space in the presented SSP formulation can often be reduced.

In the last SSP formulation in $2n+1$ states there is no real choice (there is only one available action) which allows the reduction of the state space. In this complementary section we are going to show that given an SSP problem with $N = r + s$ states, in which in $r$ states there is only one available action, we can ``safely'' reduce the number of states to $s$. More precisely, we will prove that we can construct another SSP problem with only $s$ states which is ``compatible'' with the original one in the sense that there is a one-to-one correspondence between the policies of the reduced and the original problems, and the value functions of the policies (restricted to the remaining $s$ states) are the same in both problems. Hence, finding an optimal policy for the reduced problem is equivalent to solving the original one. The computational complexity of the construction is $O(r^3+r^2sm + s^2rm)$, where $m$ denotes the maximum number of allowed actions per state. We will often omit $L$, the binary size of the input or the desired precision of the computations.

\subsubsection{Assumptions}
We will apply immediate-cost functions of the form $g: \mathbb{S} \times \mathbb{U} \to \mathbb{R}$. If we have a cost function that also depends on the arrival state (like in the refined variant above), we can redefine it using the expected cost per stage,
\begin{equation}
{g}(i,u) = \sum_{j = 1}^{N} p(j\,|\,i,u) \, \tilde{g}(i,u,j),
\end{equation}
which would not affect the outcome of the optimization \cite{Bertsekas2007}. The new cost function can be computed using $O(N^2m) = O(r^2m + rsm + s^2m)$ operations.

We will call the states in which there is only one available action as ``non-decision'' states, while the other states will be called ``decision'' states. By convention, we classify the target state, $\tau$, as a decision state. We assume, without loss of generality, that the (indices of the) non-decision states are $1,\dots, r$ and the decision states are $r+1,\dots,r+s$. Finally, we also assume that there exists at least one proper control policy.

\subsubsection{Constructing the Reduced SSP Problem}
Notice that the transition-matrix of the Markov chain induced by (any) control policy $\mu$ of the original SSP problem looks like
\begin{equation}
P_{\mu} = \begin{bmatrix}
\hspace{1mm}R_0 & R_{\mu}\hspace*{1mm}\\
\hspace{1mm}Q_0 & Q_{\mu}\hspace*{1mm}
\end{bmatrix},
\end{equation}
where $R_0 \in \mathbb{R}^{r \times r}$ describes the transition-probabilities between the non-decision states; $Q_0 \in \mathbb{R}^{s \times r}$ contains the transitions from the non-decision states to the decision states; $Q_{\mu} \in \mathbb{R}^{s \times s}$ describes the transitions between the decision states and, finally, $R_{\mu} \in \mathbb{R}^{r \times s}$ contains the transitions from the decision states to the non-decision states. Note that $R_0$ and $Q_0$ do not depend on the policy, since they correspond to non-decision states.

In the reduced problem we will only keep the $s$ decision states and remove the $r$ non-decision states. We will redefine the transition-probabilities between the decision states as if we would ``simulate'' the progress of the system through the non-decision states until we finally arrive at a decision state. In order to calculate the probabilities of arriving at specific decision states if we started in specific non-decision states, we can define a new Markov chain
\begin{equation}
P_{0} = \begin{bmatrix}
\hspace{1mm}R_0 & 0\hspace*{1mm}\\
\hspace{1mm}Q_0 & I\hspace*{1mm}
\end{bmatrix},
\end{equation}
where $0$ is a $r \times s$ zero matrix and $I$ is an $s \times s$ identity matrix. We can interpret this matrix as if we would replace each decision state by an absorbing state. We assumed that there is at least one proper policy and we know that $R_0$ and $Q_0$ are the same for all policies as well as the target state is a decision state, therefore, $R_0^{\hspace{0.3mm}k}$ converges to the zero matrix as $k \to \infty$, thus
\begin{equation}
\lim_{k\to\infty}P^{\hspace{0.3mm}k}_{0} = \begin{bmatrix}
\hspace{1mm}0 & 0\hspace*{1mm}\\
\hspace{1mm}Q^* & I\hspace*{1mm}
\end{bmatrix},
\end{equation}
where $Q^*$ contains the arrival distributions to the decision states if we started in one of the non-decision states. More precisely, $Q^*_{ij}$ is the probability of arriving at (decision) state $i$ if we start at (non-decision) state $j$. It is known \cite{Kemeny1960} that these probabilities can be calculated using the {\em fundamental matrix} of the Markov chain,
$F = (I - R_0)^{-1}$. More precisely,
\begin{equation}
Q^*=Q_0 F = Q_0 (I - R_0)^{-1},
\end{equation}
and the computation requires a matrix inversion and a matrix multiplication. If we use classical methods, $Q^*$ can be calculated in $O(r^3 + r^2 s)$ (the method of Coppersmith and Winograd \cite{Coppersmith1990} could also be applied). 
Using $Q^*$ the transition matrix of 
$\mu$ in the reduced problem should be
\begin{equation}
\widehat{P}_{\mu} = Q_{\mu} + Q^* R_{\mu}.
\end{equation}
This matrix encodes the idea that if we arrive at a non-decision state, we simulate the progress of the system until we arrive at a decision state. Fortunately, we do not have to compute it for all possible policies, we only need to define the transition-probabilities accordingly:
\begin{equation}
\widehat{p}(j \,|\,i, u) \triangleq p(j \,|\,i, u) + \sum_{k=1}^{r}{p(k \,|\,i, u)Q^*_{jk}}
\end{equation}
for all states $i, j > r$ and action $u \in \mathcal{U}(i)$. Note that states $i$ and $j$ are decision states (their indices are larger than $r$). %Consequently, 
Thus, computing the new transition-probability function can be accomplished using $O(s^2 r m)$ operations.

We should also modify the immediate-cost function, in order to include the expected costs of those stages that we spend in the non-decision states, as well. It is known that the fundamental matrix contains information about the expected absorbing times. 
More precisely, $F_{jk}$ is the expected time spent in (non-decision) state $j$ before arriving at a (decision) state (absorption), if the process started in (non-decision) state $k$ \cite{Kemeny1960}. Therefore, 
\begin{equation}
\widehat{g}(i, u) \triangleq g(i, u) + \sum_{k=1}^{r}{p(k \,|\,i, u)\sum_{j=1}^{r} F_{jk}\,g(j)},
\end{equation}
for all $i > r$ and $u \in \mathcal{U}(i)$, where we did not denote the dependence of the cost function on the actions for non-decision states, since there is only one available action in each such state. Thus, $g(j) \triangleq g(j, u)$, where $u$ denotes the only available action in state $j$. Computing the new cost-function needs $O(r^2 s m)$ operations, if we already have the fundamental matrix.

We have only removed non-decision states, in which there is only one allowed action, consequently, it is trivial to extend a policy of the reduced problem to a policy of the original one, and there is a {\em bijection} between such policies. Since we defined the transition-probabilities and the immediate-cost function in a way that it mimics the behavior of the original problem, solving the reduced problem is equivalent to solving the original one. Summing all computations together, we can conclude that the time complexity of the construction is $O((r^3 + r^2 s m + s^2 r m)L)$, where $L$ is the binary size or precision.

\subsubsection{Reducing the SSP formulation of Max-PageRank} Applying this result to the refined SSP formulation of Max-PageRank, we can reduce the number of states to $d$ (without $\tau$) by constructing another SSP problem as demonstrated above. It can be summarized as
\vspace{1mm}
\begin{lemma}
The \textsc{Max-PageRank} problem with a digraph having $n$ nodes and $d$ fragile links can be reduced to an SSP problem with only $d$ states (plus the termination state) by using $O((n^3+d^{\hspace{0.3mm}2}n +n^2d)L)$ operations.
\end{lemma}

%\medskip
\section{PageRank Iteration} In the previous sections we saw how to reformulate efficiently the Max-PageRank problem as an SSP problem. This SSP formulation could then be further refomulated as an LP problem, which type of problems are known to be solvable in polynomial time, for example, by interior point methods.

Now, we will provide an alternative solution to the Max-PageRank problem. We will build on the previous SSP formulation, but instead of using an LP-based solution, we will define a simple iterative algorithm that in each step updates the configuration of the fragile links in a greedy way. Yet, as we will see, this method is efficient in many sense. For simplicity, we will only consider the case without damping ($c=0$) and we will apply the assumption:
\begin{enumerate}
\vspace{1mm}
\item[(AB)] {\em Bounded Reachability Assumption}\,: We assume that the target node, $v$, can be reached from all nodes of the graph via a bounded length path of fixed edges. In other words, there is a universal constant $\kappa$ such that node $v$ can be reached from all nodes by taking at most $\kappa$ fixed edges. The fact that $\kappa$ is universal means that it does not depend on the particular problem instance.
\vspace{1mm}
\end{enumerate}

The algorithm starts with a configuration in which each fragile link is activated. In iteration $k$ it computes the expected first hitting time to $v$ if we start in $i$ and use the current configuration, that is it calculates
\begin{equation}
H_k(i) \triangleq \mathbb{E} \left[ \, \inf  \left\{\, t \geq 1 : X_t = v\, \right\} \,|\, X_0 = i \,\right],
\end{equation}
for all nodes $i$, where the transition matrix of the Markov chain $(X_0, X_1, \dots)$ is $P_k$ defined by equation (\ref{rndw}) using the adjacency matrix corresponding to the fragile link configuration in iteration $k$. Then, the configuration is updated in a greedy way: a fragile link from node $i$ to node $j$ is activated if and only if $H_k(i) \geq H_k(j)+1$. The algorithm terminates if the configuration cannot be improved by this way. We call this method the {\em PageRank Iteration} (PRI) algorithm. The pseudo-code of PRI can be found below.

{\renewcommand{\baselinestretch}{1.2}
\begin{table}[h]
\begin{center}
\small
\vspace{1mm}
\hspace*{-2mm}
\begin{tabular}{|rllll|}
\multicolumn{5}{c}{\textsc{The PageRank Iteration Algorithm}\vspace{1mm}}\\
\hline
\multicolumn{5}{|l|}{{\em Input:} \hspace{2mm} A digraph $\mathcal{G} = \left( \mathcal{V}, \mathcal{E} \right)$, a node $v \in \mathcal{V}$ and a set of fragile links $\mathcal{F} \subseteq \mathcal{E}$.}\\
\hline
\hline
1. & \multicolumn{2}{l}{$k := 0$} & \multicolumn{2}{l|}{\% initialize the iteration counter}\\
2. & \multicolumn{2}{l}{$F_0 :=\mathcal{F}$}  & \multicolumn{2}{l|}{\% initialize the starting configuration}\\
3. & \multicolumn{2}{l}{{\em Repeat}} & \multicolumn{2}{l|}{\% iterative evaluation and improvement} \\
4. & \hspace{4mm} & \multicolumn{1}{l}{$H_k := \1^{\!\mathrm{T}}(I - Q_{k})^{-1}$} &  \multicolumn{2}{l|}{\% compute the mean hitting times to $v$}\\
5. & \hspace{4mm} & \multicolumn{2}{l}{$F_{k+1} := \left\{ (i,j) \in \mathcal{F} : H_k(i) \geq H_k(j)+1 \right\}$} &  \% improve the configuration\\
6. & & \multicolumn{1}{l}{$k := k+1$} & \multicolumn{2}{l|}{\% increase the iteration counter}\\
\hspace{0.5mm} 7. & \multicolumn{2}{l}{{\em Until} \, $F_{k-1} \neq F_{k}$} & \multicolumn{2}{l|}{\% until no more improvements are possible}\\
\hline
\hline
\multicolumn{5}{|l|}{{\em Output:} \hspace{0mm} $1/H_k(v)$, the Max-PageRank of $v$, and $F_k$, an optimal configuration.}\\
\hline
\end{tabular}
\vspace{-3mm}
\end{center}
\end{table}
\renewcommand{\baselinestretch}{1}}

Note that the expected first hitting times can be calculated by a system of linear equations \cite{Levin2009}. In our case, the vector of hitting times, $H_k$, is
\begin{equation}
H_k = \1^{\!\mathrm{T}}(I - Q_{k})^{-1},
\end{equation}
where $Q_k$ is obtained from $P_k$ by setting to zero the row corresponding to node $v$, namely, $Q_k = diag\,(\1-e_v)\,P_k$, where $e_v$ is the $v$-th $n$ dimensional canonical basis vector. To see why this is true, recall the trick of Section \ref{simplef}, when we split node $v$ into a starting node and an absorbing target node. Then, the expected hitting times of the target state can be calculated by the fundamental matrix \cite{Kemeny1960}. If $v$ can be reached from all nodes, then $I - Q_k$ is guaranteed to be invertible. Note that $H_k(v) = \varphi_k(v)$, where $\varphi_k(v)$ is the expected first return time to $v$ under the configuration in iteration $k$, therefore, the PageRank of $v$ in the $k$-th iteration is $\prv_k(v) = 1/H_k(v)$.

\begin{theorem}
\textsc{PageRank Iteration} has the following properties:
\vspace{-1.5mm}
{\em 
\begin{enumerate}[(I)]
\item {\em Assuming {\em(AD)} and {\em(AR)}, the algorithm always terminates in a finite number of iterations and the final configuration is optimal.}\vspace{-1.5mm}
\item {\em Assuming {\em(AB)}, it finds an optimal solution in polynomial time.}
\end{enumerate}
}
\end{theorem}

\begin{proof}{}
Part I: We can notice that this algorithm is almost the {\em policy iteration} (PI) method, in case we apply a formulation similar to the previously presented simple SSP formulation. However, it does not check every possible action in each state. It optimizes each fragile link separately, but as the refined SSP formulation demonstrates, we are allowed to do so. Consequently, PRI is the policy iteration algorithm of the refined SSP formulation. However, by exploiting the special structure of the auxiliary states corresponding to the fragile links, we do not have to include them explicitly. For all allowed policies $\mu$ (for all configurations of fragile links) we have
\begin{equation}
J^{\mu}(f_{ij}) =
\left\{
\begin{array}{ll}
J^{\mu}(i) \hspace{3mm}& \mbox{if}\hspace{2mm} \mu(f_{ij}) = \mbox{``off''},\\
J^{\mu}(j)+1 \hspace{3mm}& \mbox{if}\hspace{2mm} \mu(f_{ij}) = \mbox{``on''},\\
\end{array}
\right.
\end{equation}
for all auxiliary states $f_{ij}$ corresponding to a fragile link. Thus, we do not have to store the value of these states, since they can be calculated if needed. 

Notice that $J^{\mu_k}(i) = H_k(i)$, where $\mu_k$ is the policy corresponding to the configuration in iteration $k$. Thus, calculating $H_k$ is the {\em policy evaluation} step of PI, while computing $F_{k+1}$ is the {\em policy improvement} step. Since PRI is a PI algorithm, it follows that it always terminates finitely and finds an optimal solution \cite{Bertsekas1996} if we start with a {\em proper} policy and under assumptions (A1) and (A2). Recall that the initial policy is defined by the full configuration, $F_0 = \mathcal{F}$ and that we assumed (AR), that is node $v$ can be reached from all nodes for at least one configuration which means that the corresponding policy is proper. If this holds for an arbitrary configuration, it must also hold for the full configuration, therefore, the initial policy is always proper under (AR). Assumption (A1) immediately follows from (AR) and assumption (A2) follows from the fact that if the policy is improper, we must take infinitely often fixed or activated fragile links with probability one. 
Since each of these edges has unit cost, the total cost is infinite for at least one state.

Part II: First, note that assumption (AB) implies (AR) and (AD), therefore, we know from Part I that PRI terminates in finite steps with an optimal solution. Calculating the mean first hitting times, $H_k$, basically requires a matrix inversion, therefore, it can be done in $O(n^3)$. In order to update the configuration and obtain $F_{k+1}$, we need to consider each fragile link individually, hence, it can be computed in $O(d)$. Consequently, the problem of whether PRI runs in polynomial time depends only on the number of iterations required to reach an optimal configuration.

Since we assumed (AB), there is a universal constant $\kappa$ such that for all nodes of the graph there is a directed path of fixed edges from this node to node $v$ which path has at most $\kappa$ edges. These paths contain fixed (not fragile) edges, therefore, even if all fragile links are deactivated, node $v$ can still be reached with positive probability from all nodes. Consequently, all policies are proper (APP). It is easy to see that we can partition the state space to subsequent classes of states $S_1, \dots, S_r$, where $r \leq \kappa$, by allocating node $i$ to class $S_q$ if and only if the smallest length path of fixed edges that leads to node $v$ has length $q$. This partition satisfies the required property described in Section \ref{MDP-complexity}. Because PRI is a PI variant, PRI terminates with an optimal solution in iterations bounded by a polynomial in $L$ and $\eta^{-2\kappa}$. Since $\eta = 1/m$, where $m \leq n$ is the maximum out-degree in the graph, $\eta^{-2\kappa} = O(n^{2\kappa})$, therefore, PRI runs in polynomial time. 
\end{proof}

Though, for the sake of concision, we only presented PRI for the problem without damping and personalization, it is easy to modify the algorithm for the other case, as well. Moreover, since if we apply damping each node can be reached from all other nodes by a constant number of edges, namely via the teleportation state, assumption (AB) is automatically satisfied. Then, the smallest transition probability of the associated SSP problem may be determined by the damping constant and the personalization vector, however, this can be arbitrary small. On the other hand, if the damping constant and the personalization vector are {\em fixed}, not part of the input, we do not have this problem and hence PRI finds an optimal solution in polynomial time. 

%\medskip
\section{PageRank Optimization with Constraints} In this section we are going to investigate a variant of the PageRank optimization problem in which there are mutually exclusive constraints between the fragile links. More precisely, we will consider the case in which we are given a set of fragile link pairs, $\mathcal{C} \subseteq \mathcal{F} \times  \mathcal{F}$, that cannot be activated simultaneously. The resulting problem is summarized below.

\begin{table}[h]
\begin{center}
\hspace*{-2mm}
\footnotesize
\begin{tabular}{|l l|} \hline
\multicolumn{2}{|l|}{\textsc{The Max-PageRank Problem under Exclusive Constraints}\rule[4mm]{0pt}{0pt}}\\
\rule[4mm]{0pt}{0pt}{\em Instance:} & A digraph $\mathcal{G} = \left( \mathcal{V}, \mathcal{E} \right)$, a node $v \in \mathcal{V}$, a set of controllable edges $\mathcal{F} \subseteq \mathcal{E}$\\
& and a set $\mathcal{C} \subseteq \mathcal{F} \times  \mathcal{F}$ of those edge-pairs that cannot be activated together.\\
 & A damping constant $c \in (0,1)$ and a stochastic personalization vector $z$.\\
{\em Task:} & Compute the maximum possible PageRank of $v$ by activating edges in $\mathcal{F}$ \\
& and provide a configuration of edges in $\mathcal{F}$ for which the maximum is taken. \vspace{-2.5mm} \\
&\\ \hline
\end{tabular}
\vspace{-5mm}
\end{center}
\label{MaxPageRank}
\end{table}

We will show that the Max-PageRank problem under exclusive constraints is already {\em NP-hard}, more precisely, we will show 
that the decision version of it is NP-complete. In the decision version, one is given a real 
number $p \in (0,1)$ and is asked whether there is a fragile link configuration such that the PageRank of a given node $v$ is larger or equal to $p$. 

\begin{theorem}
The decision version of the \textsc{Max-PageRank Problem under Exclusive Constraints} is NP-complete.
\end{theorem}

\begin{proof}{}
The problem is in NP because given a solution (viz., a configuration), it is easy to {\em verify} in polynomial time, e.g., via a simple matrix inversion, cf.\ equation (\ref{PR-inv}), whether the corresponding PageRank is larger or equal to $p$.

We now reduce the $3$SAT problem, whose NP-completeness is well known \cite{Garey1990}, to this problem. In an instance of the $3$SAT problem, we are given a Boolean formula containing $m$ disjunctive {\em clauses} of three {\em literals} that can be a variable or its negation, and one is asked whether there is a truth assignment to the variables so that the formula (or equivalently: each clause) is satisfied. Suppose now we are given an instance of $3$SAT. We will  construct an instance of the Max-PageRank problem under exclusive constraints that solves this particular instance of $3$SAT. 

We construct a graph having $m+2$ nodes in the following way: we first put a node $s$ and a node $t.$  Figure it as a source node and a sink node respectively. Each clause in the given $3$SAT instance can be written as $y_{j,1}\vee y_{j,2} \vee y_{j,3},$  $1\leq j \leq m,$ where $y_{j,l}$ is a variable or its negation.  For each such clause, we add a node $v_j$ between $s$ and $t,$ we put an edge from $v_j$ to itself (a self-loop), we put an edge from $s$ to $v_j,$ and we put three edges between $v_j$ and $t,$ labeled respectively with $y_{j,1},y_{j,2},$ and $y_{j,3}.$  We finally add an edge from $t$ to $s$.  We now define the set of exclusive constraints, $\mathcal{C}$, which concludes the reduction. For all pairs $(y_{j,l},y_{j',l'})$ such that $y_{j,l}=\bar y_{j',l'}$ (i.e., $y_{j,l}$ is a variable and $\bar y_{j',l'}$ is its negation, or conversely), we forbid the corresponding pair of edges. Also, for all pairs of edges $(y_{j,l},y_{j,l'})$ corresponding to a same clause node, we forbid the corresponding pair. This reduction is suitable, since the sizes of the graph and $\mathcal{C}$ are {\em polynomial} in the size of the $3$SAT instance.

We claim that for $c$ small enough, say $c=1/(100m)$, it is possible to obtain an expected return time from $t$ to itself which is smaller than $77$ if and only if the instance of $3$SAT is satisfiable. The reason for that is easy to understand with $c=0:$ if the instance is not satisfiable, there is a node $v_j$ with no edge from it to $t$. In that case, the graph is not strongly connected, and the expected return time from $t$ to itself is infinite. Now, if the instance is satisfiable, let us consider a particular satisfiable assignment. We activate all edges which correspond to a literal which is true and, if necessary, we deactivate some edges so that for all clause nodes, there is exactly one leaving edge to $t.$  This graph, which is clearly satisfiable, is strongly connected, and so the expected return time to $t$ is finite.

Now if $c\neq 0$ is small enough, one can still show by continuity that the expected return time is much larger if some clause node does not have an outgoing edge to $t$. To see this, let us first suppose that the instance is not satisfiable, and thus that a clause node (say, $v_1$), has no leaving edge.  Then, for all $l\geq 3,$ we describe a path of length $l$ from $t$ to itself: this path passes through $s,$ and then remains during $l-2$ steps in $v_1,$ and then jumps to $t$ (with a zapping).  This path has probability $(1-c)\frac{1}{m}(1-c)^{l-2}c$. 
Therefore, the expected return time is larger than
\begin{equation}
E_1 \, \geq \,\sum_{l = 3}^{\infty}{l p(l)}\, \geq\, \frac c m \sum_{l = 3}^{\infty}{l(1-c)^{l-1}}\, \geq\, \frac c m \left [ c^{-2}-3 \right ]\,\geq\, 99,
\end{equation}
where we assumed that $c=1/(100m)$ and the personalization vector is $z = (1/n)\, \1$. 
Note that $c$ and $z$ are part of the input, thus they can be determined.

Consider now a satisfiable instance, and build a corresponding graph so that for all clause nodes, there is exactly one leaving edge. It appears that the expected return time from $t$ to itself satisfies $E_2\leq 77.$ To see this, one can aggregate all the clause nodes in one macro-node, and then define a Markov chain on three nodes that allows us to derive a bound on the expected return time from $v_t $ to itself.  This bound does not depend on $m$ because one can approximate the probabilities $m/(m+2)$ and $1/(m+2)$ that occur in the auxiliary Markov chain by one so that the bound remains true. Then, by bounding $c$ with $1/8>1/(100 m),$ one gets an upper bound on the expected return time. For the sake of conciseness, we skip the details of the calculations.
To conclude the proof, it is possible to find an edge assignment in the graph so that the PageRank is greater than $p=1/77$ if and only if the original instance of 3SAT is satisfiable.
\end{proof}

We have tried to keep the NP-hardness proof as short as possible. Several variants are possible. In the above construction, each clause node has three parallel edges linking it to the node $t.$  This might seem not elegant, but it is not difficult to get rid of them by adding auxiliary nodes.  Also, it is not difficult to get rid of the self-loops by adding auxiliary nodes. Finally we have not tried to optimize the factor $c=1/(100m),$ nor the bound on $E_2.$ An interesting question is whether a reduction is possible if the damping factor $c$ and the personalization vector $z$ cannot depend on the instance.

%\medskip
\section{Conclusions}
The task of ordering the nodes of a directed graph according to their {\em importance} arises in many applications from the problem of ranking the results of web-searches to bibliometrics and ecosystems. A promising and popular way to define such an ordering is to use the {\em PageRank} method \cite{Brin1998} and associate the importance of a node with the weight of the node with respect to the stationary distribution of a uniform random walk. The problem of optimizing the PageRank of a given node by changing some of the edges caused a lot of recent interest \cite{Olsen2009,Avrachenkov2006,DeKerchove2008,Ishii2009,fercoq:ergodic}. We considered the general problem of finding the extremal values of the PageRank a given node can have in the case we are allowed to control (activate or deactivate) some of the edges from a given {\em arbitrary} subset of edges, which we referred to as {\em fragile links}. 

Our main contribution is that we proved that this general problem can be solved optimally in {\em polynomial time} under the Turing model of computation, even if the {\em damping constant} and the {\em personalization vector} are part of the input and independently of the way the {\em dangling nodes} are handled. The proof is based on reformulating the problem as a {\em stochastic shortest path} problem (a special Markov decision process) and it results in a linear programming formulation that can then be solved by standard techniques.

This solution is weakly polynomial in general, however, if the damping constant and the personalization vector can be represented with bits polynomial in the number of nodes, it becomes strongly polynomial.

We do not need to assume that the graph is simple, namely, it can have multiple edges (and self-loops). This allows the generalization of our results to {\em weighted graphs}, in case the weights are positive integers or rationals. 

Based on the observation that in some of the states of the reformulated SSP problem there is only one available action (thus, we do not have a real choice in them), we showed that the number of states (and therefore the needed computation to solve the problem) could be further reduced. 

We also suggested an alternative greedy solution, called the {\em PageRank Iteration} (PRI) algorithm, which had appealing properties. We analyzed PRI for the Max-PageRank problem {\em without damping} and showed that it can find an optimal solution in finite steps and it runs in {\em polynomial time}, under the {\em bounded reachability} assumption. This latter assumption is always satisfied if we consider the problem {\em with damping} which indicates that PRI always finds an optimal solution in polynomial time for such problems, in case the damping constant and the personalization vector are {\em fixed}. 

Finally, we also showed that slight modifications of the problem, as for instance adding mutual exclusive constraints between the activation of several fragile links, may turn the problem {\em NP-hard}. We conjecture that several other slightly modified variants of the problem are also NP-hard, e.g., the Max-PageRank problem with restrictions on the number of fragile links that can be simultaneously activated. We left their analysis for further work.

%\medskip
\section*{Acknowledgments}
The authors are grateful for the valuable discussions to Paul Van Dooren, Yurii Nesterov, Cristobald de Kerchove, Vincent Traag and Tzvetan Ivanov. B.\ Cs\'aji is an ARC DECRA fellow and R.\ Jungers is an F.R.S.-FNRS fellow.



%% References


\bibliographystyle{elsarticle-num}
\bibliography{dam-pagerank}






% \end{thebibliography}

\end{document}